# A Novel Cross Layer Scheme for Multi-Channel Hybrid Cognitive Ad-hoc Networks

Yingji Zhong [1,2]    Sana Ullah [1]    Kyung Sup Kwak [1]
1. UWB Communication Research Center, Inha University, Incheon, Korea
2. School of Information Science and Engineering, Shandong University, Jinan, Shandong, China
zhongyingji32@sdu.edu.cn, sanajcs@hotmail.com, kskwak@inha.ac.kr

## Abstract

A special scenario of the topology in the hybrid Cognitive Ad-hoc networks is studied and a novel cross layer scheme is proposed in this paper. The proposed scheme integrated the attributes both of the new performance evaluation machine check time metric and the topology space in special scenario. The topology and power consumption of each node can all be optimized due to the minimum link occupation with the help of this scheme. Simulation results show that the novel scheme can give schedule guarantee to the multi-channel networks in the variable node loads and transmission powers, and make the node stable to support multi-hops at the same time.

## I. Introduction

The hybrid Ad-hoc networks [1,2] give a novel infrastructure which combines cellular network with Ad-hoc mechanism, it should be a trade-off between these components. We believe that the topology space analysis in the special scenario should be beneficial for the proposal of a novel scheme. In the ad-hoc cognitive radio, the power of each terminal is suppressed to minimize the interference toward the primary system and the area of communication can expand by using the multi-hop networks. In this paper, based on the special scenario of the network topology we explore the relationship between attributes of the topology space and the topology scenario. To this end, we propose a new performance evaluation metric and the novel scheme to effectively utilize location marking information and then address the performance issues. Each node makes decision independently in the hybrid network. It may cause inconsistence and confusion when two nearby nodes adjust their topology simultaneously [3,4]. Thus it is required that both the channel adjustment and power adjustment should guarantee the exclusiveness of the cross layer adjustment in node's interference area. Under this premise, each node may locally make schedule decisions without considering the disturbance of neighboring nodes. Consequently with this arrangement a mobile node can be formed, which can operate in both infrastructure and Ad-hoc mode. Communication between mobile nodes in different cells is accomplished through their respective domain in a similar fashion as conventional cellular networks.

The rest of this paper is organized as follows. In Section II, we give the evaluation models and make the dimensionality analysis of the topology. In Section III, we propose performance evaluation metric and the novel scheme. In Section IV, we evaluate the performance of the proposed scheme and analyze the improvement of the guarantee via simulation. Finally we give the conclusion in Section V.

## II. Dimensionality Analysis of Topology and Evaluation Models

The evaluation model is the multiple-cell environments with seven cells as shown in Fig.1, in which the Mobile Hosts (MHs) are in point wise uniformity. Analysis is based on two-dimension scenario, that is to say, the MHs and the base stations are on a cross. MHs are placed at every *D* distance unit from base stations and define intervals [*nD*, (*n*+1)*D*] of length *D* on the Cross $C_s$. One scheme of this kind of topology scenario is shown in Fig.1. The base station of Cell 1 is on the middle of $C_s$. Assume $C_s(N)$ is locally compact space. Let $\cup A=N$, for each $n \in N$, define $Fn=\{1,2,\ldots,n\}$ and $Un=\{\{A\}\cup(A/Fn):A\in_A\}\cup\{\{x\}:x\in Fn\}$ the n *Un* is the open covering of $C_s(N)$. For each x $\in$ $C_s(N)$, when $x\in Fn$, $st(x,Un)=\{x\}$, and $x\in A$, $st(x,Un)=\{x\}\cup(x/Fn)$. So $\{U_n\}$ is the development of $C_s(N)$. Thus, $C_s(N)$ is the developable space of locally compact. Let K is the compact subspace of $C_s(N)$, since A is the closed discrete subspace of $C_s(N)$, so K$\cap$A is the finite set, K is countable set of $C_s(N)$ and K can be metric, it is obvious that K has countable neighborhood basis in $C_s(N)$. $C_s(N)$ is a dividable space, if it has point-countable basis, then A, the subspace of $C_s(N)$ has countable basis, this is contrary to reality that A is the uncountable closed discrete subspace of $C_s(N)$. So $C_s(N)$ has no point-countable basis.

It can be seen that the cross topology belongs to the combination of two Gillman-Jerison spaces [5]. However, a distinct difference in this case is that MHs themselves can also function as active mobile nodes. These nodes under group-oriented operation are capable of initiating communications not only with their mobile nodes, but also with others. The attributes of Gillman-Jerison Space are potentially worthy for the novel scheme proposal.

## III. Performance Evaluation Metric and Cross Layer Scheme

The physical layer sub-problem addresses the transmission interference among nearby nodes. Wu [6] has given the new metric called *ECATM* (Equivalent Channel Air Time Metric) to reflect also spatial reusability characteristic. As for the Ad-hoc networks, It resides between MAC and network layer and aims to improve the network throughput by coordinating the transmission power, channel assignment and route selection among multiple nodes in a distributed way. According to the intuition obtained from the above, we propose a new performance evaluation Machine Check Time Metric (*MTM*) defined as:

$$MTM_i = \sum_c \sum_l r_l^c RT_l^c F^c Q_l / CF^c \quad (1)$$

where *c* represent the available channels and *l* denotes co-channel links that lie in the interference range of a specific node *i*, $RT_l^c$ denotes the round trip factor of the corresponding channel, $Q_l$ specifies the quality of the link, $CF^c$ denotes the channel reuse factor on channel *c*. *MTM* represents the aggregated equivalent channel air time for potential candidates and acts as an indicator for the network performance.

When the end-to-end traffic can be split in the cross topology, the number of the routes between source and destination should be more than one. That is to say, the flow going through the route is no longer an integer and the traffic demands can be split [5].

The topology construction is performed during the network initialization phase when no user traffic is present in the multi-channel network. It is not good enough only considering channel assignment. To fully reduce the co-channel interference and consequently achieve higher gains of network performance, the topology control and routing should be jointly considered to exploit not only channel diversity but also spatial reusability. The operation should be terminated when the transmission power reaches to maximum. In this scheme, the topology and power consumption of each node can be optimized due to the minimum link occupation. The power update is the best response of link player given the tax rate and assessment of others' action. As for the tax rates converge, the power allocation equilibrium strikes a balance between minimizing interference and maximizing rate. The algorithm of the novel scheme is shown below.

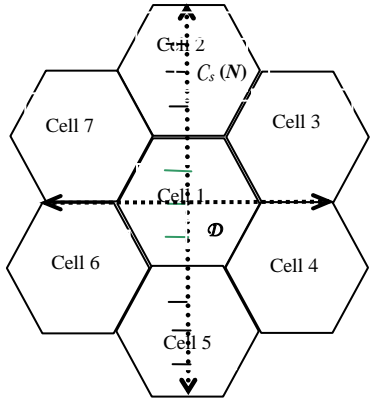

Figure. 1. One scheme of cross topology scenario

```
Init()
{
  for each available channel  c ∈ {c_1, c_2, ..., c_L}
  if  t=0
   j_prio=Prand(j)W(j);
  end if
  analyze the contention of links on channel c in two hop range;
  if i is bound to nodes of neighboring cluster then
      Assign i, the channel assignment from its neighbor assignment;
    else
      Calculate new assignment for (v,i);
  end if
  calculate MTM value on channel c and corresponding priority for each group;
  if no channel overloaded
      return;
  end if
  Clusterhead(i)
   {
    for  j ∈ Ni
      if Prand(Ni)< W(j) and priority of t is not Φ
        Recover Ni;
      end if
   }
   if feasible
     then select adjustment candidate with minimal MTM value and begin negotiation;
   end if
  end
}
```

## IV. Simulations and Discussion

In this section, we present simulations to illustrate our theoretical results. We assume that the terrain model is a 100$km$×100$km$ square area with seven cells in it, on which 50000 MHs are pseudo-randomly moving along the cluster cells. All the MHs are presented by $\{u_i\}$ $i \in$ [1, 50000], and all the links between MHs are bi-directional. Each cell has a base station with omni-directional antenna at the center point and its radius is 15$km$. Each base station has 1024 available data channels. We use the modified DSR protocol with location information[7] as the routing protocol for the Ad-hoc mode. Assume the power consumption is based on the distance from the transmitting MHs to the base stations. As for handoff mechanism, hard handoff was used in the evaluation model and connectivity is considered under Poisson Boolean Model in this kind of sparse network. Employing the proposed scheme, the traffic requirement and the maximal amount of the permitted hops are examined in different load and transmitting power of the nodes with or without *MTM*, shown separately in Fig.2, Fig.3.

Fig.2 shows the different values of the traffic requirement along with the load of the nodes with or without *MTM*. We can see that the traffic requirement depend deeply on the load when use no *MTM* but released by *MTM* from it. The maximal optimization is 8.31% when the load reached to 200Mb/s. What is more, the addressing ratio of success in the condition of the unchanged parameters and external information can also be increased. The maximal amounts of the permitted hops in different loads are shown in Fig.3. The value is fixed when the load is 0, that is to say, the default value of the hops is 1 when the nodes have no load. The network can tolerate more hops to support reliable transportation with the help of *MTM*, it can make the network more stable and can support more hops. The maximal optimization is about 11.19% and the merits of the proposed scheme are obvious.

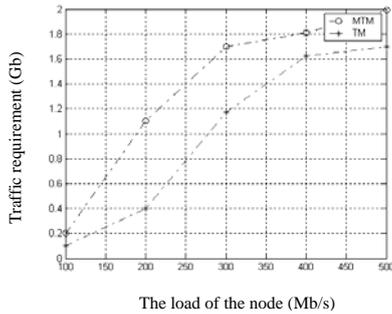

Figure.2. The traffic requirements for different loads

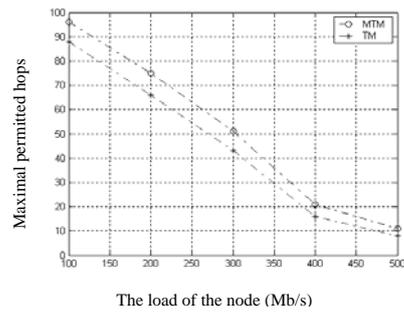

Figure.3. The maximal amount of the permitted hops in different loads

## V. Conclusions

Based on the analysis of the topology space and cross layer constraints, we induced the special scenario of the topology in the multi-channel hybrid Ad-hoc networks. With the help of the cross layer constraints, we proposed a new performance evaluation time metric and a novel cross layer scheme in this paper. The proposed scheme integrated the attributes both of the new metric and the topology space in the special scenario. The topology and power consumption of each node can all be optimized due to the minimum link occupation with the help of the scheme. Simulation results show that the novel scheme can give power control guarantee to the multi-channel networks in the variable node loads and transmission powers.

## Acknowledgement

This research was supported by the MKE(Ministry of Knowledge Economy), Korea, under the ITRC(Information Technology Research Center) support program supervised by the IITA(Institute of Information Technology Assessment) (IITA-2008-C1090-0801-0019); the Natural Science Foundation in Shandong Province(Q2007G01) and the Outstanding Youth Scientist Awards Foundation in Shandong Province(2006BS01009).